\documentclass{PoS}
\usepackage{amsmath,amssymb}
\usepackage{dsfont}
\usepackage{slashed}
\usepackage{verbatim}

\newcommand{\D}{\mathcal{D}}

\newcommand{\be}{\begin{equation}}
\newcommand{\ee}{\end{equation}}
\newcommand{\Tr}{\textmd{Tr}}

\newcommand{\ZZ}{\mathcal{Z}}
\newcommand{\expv}[1]{\left \langle #1 \right \rangle}
\newcommand{\TAU}{\tau_f}

\def\ZT{Z_{\rm T}}
\def\zetaT{\zeta_{\rm T}}

\def\smn{{\sigma_{\mu\nu}}}

\def\chiF{\chi}
\def\chiM{\xi}

\def\MSBar{\overline{\rm MS}}

\newcommand{\Wuppertal}{Bergische Universit\"at Wuppertal, Theoretical
Physics, 42119 Wuppertal, Germany.}
\newcommand{\Budapest}{E\"otv\"os University, Theoretical Physics,
P\'azm\'any P. S 1/A, H-1117, Budapest, Hungary.}
\newcommand{\Regensburg}{Institute for Theoretical Physics,
Universit\"at Regensburg, D-93040 Regensburg, Germany.}
\newcommand{\Julich}{J\"ulich Supercomputing Centre, Forschungszentrum
J\"ulich, D-52425 J\"ulich, Germany.}
\newcommand{\Cyprus}{Department of Physics, University of Cyprus, Nicosia CY-1678, Cyprus.}

\title{Thermodynamic properties of QCD in external magnetic fields}

\ShortTitle{Thermodynamic properties of QCD in external magnetic fields}

\author{
G.~S.~Bali$^1$, F.~Bruckmann\footnote{Speaker, presentation: {\it The QCD transition in external magnetic fields.}}$^{*1}$, M. Constantinou$^2$, M. Costa$^2$,  G.~Endr\H{o}di\footnote{Speaker, presentation: {\it Magnetic susceptibility of the QCD condensate at zero and at finite temperature from the lattice}.}$^{\dagger1}$,
Z.~Fodor$^3$, S.~D.~Katz$^4$, S.~Krieg$^{3,5}$, H. Panagopoulos$^2$, A.~Sch\"afer$^1$,
K.~K.~Szab\'o$^3$
}

\footnotetext[1]{\Regensburg}
\footnotetext[2]{\Cyprus}
\footnotetext[3]{\Wuppertal}
\footnotetext[4]{\Budapest}
\footnotetext[5]{\Julich}

\abstract{We consider the effect of strong external electromagnetic fields on thermodynamic observables in QCD, through lattice simulations with $1+1+1$ flavors of staggered quarks at physical quark masses. Continuum extrapolated results are presented for the light quark condensates and for their tensor polarizations, as functions of the temperature and the magnetic field. We find the light condensates to undergo inverse magnetic catalysis in the transition region, in a manner that the transition temperature decreases with growing magnetic field. We also compare the results to other approaches and lattice simulations. Furthermore, we relate the tensor polarization to the spin part of the magnetic susceptibility of the QCD vacuum, and show that this contribution is diamagnetic.}

\FullConference{Xth Quark Confinement and the Hadron Spectrum\\
         8-12 October 2012\\
         TUM Campus Garching, Munich, Germany}

\begin{document}

\section{Introduction}

Strong electromagnetic fields represent an important ingredient for the description of various strong interaction systems including non-central heavy-ion collisions, the interior of dense neutron stars (magnetars) and the cosmology of the early universe. The external magnetic field in these situations is expected to have a magnitude reaching up to the hadronic scale, such that its coupling to charged quarks and the color interaction between the latter compete with each other. This competition gives rise to an interplay between the strong and electromagnetic dynamics, producing various new phenomena. We will restrict ourselves to the idealized situation of constant external magnetic fields in thermal equilibrium. In fact, such magnetic fields probe the QCD vacuum in several aspects, by affecting its fundamental properties like chiral symmetry breaking and restoration, (de)confinement and hence the phase diagram, as well as the vacuum polarization and other quantities.

At vanishing quark density, the transition of QCD at the physical values of the quark masses is a crossover~\cite{Aoki:2006we} with transition temperatures, that may depend on the observable used for its definition. Here we focus on quark condensates, pseudo-order parameters related to the breaking of chiral symmetry, and the magnetic susceptiblities, relevant for various phenomenological applications, in external magnetic fields. The renormalization of both will be defined, numerical results are presented for zero and nonzero temperatures and compared to results from other approaches. While the most important effects in the condensates are the inverse magnetic catalysis around the transition and the transition temperature decreasing with the magnetic field, the magnetic susceptibility contains information about the spin contribution to the para- or diamagnetic nature of the QCD vacuum. In these talks we  summarize our publications~\cite{Bali:2011qj,Bali:2012zg,Bali:2012jv}, adding another plot for the continuum limit of the condensate (Fig.\ \ref{fig:IMC}, right panel) and a few new comments.

\subsection{Magnetic catalysis and inverse magnetic catalysis}

A particularly pronounced effect of the magnetic field on the QCD dynamics is the so-called {\it magnetic catalysis} (MC) mechanism -- the fact that the condensate increases with the external field $B$~\cite{Schramm:1991ex,Gusynin:1995nb} at zero temperature. This behavior has been observed in various effective theory and model calculations (for a recent review, see Ref.~\cite{Shovkovy:2012zn}), and has also been confirmed by lattice simulations in quenched theories~\cite{Buividovich:2008wf,Braguta:2010ej}, at larger than physical pion masses in $N_f=2$ QCD~\cite{D'Elia:2010nq,D'Elia:2011zu} and in the $N_f=4$ $\mathrm{SU}(2)$ theory~\cite{Ilgenfritz:2012fw}, and at physical quark masses in full $N_f=2+1$ QCD~\cite{Bali:2011qj}.

While all effective descriptions and lattice simulations agree about MC at $T=0$, the situation becomes more complicated at finite temperature. As lattice simulations at physical quark masses with results extrapolated to the continuum limit show, the condensate exhibits a non-monotonic dependence on the magnetic field around $T_c$, with a certain region, where it actually decreases with growing $B$~\cite{Bali:2011qj}. This peculiar behavior we refer to as {\it inverse magnetic catalysis} (IMC). A consequence of this non-trivial dependence $\bar\psi\psi(B,T)$ is that the transition temperature $T_c(B)$ is reduced as the magnetic field increases. As it turns out, employing physical quark masses in the simulation and extrapolating the results to the continuum limit -- as was done in our studies -- is essential, as we will discuss. It is also highly important to address the differences between the lattice QCD results and model and chiral perturbation theory ($\chi$PT) predictions,
especially since the latter methods can be used to investigate regions that are not easily accessible to lattice simulations, e.g., QCD at a non-vanishing baryon density.

\subsection{Paramagnetism and diamagnetism}

Besides the influence of the magnetic field on chiral symmetry breaking, a fundamental aspect of the coupling between the strong dynamics and $B$ is given in terms of the response of the free energy density,
\be
f=-\frac{T}{V} \log \ZZ,
\label{eq:freeenergy}
\ee
where $\ZZ$ is the partition function of the system and $V$ the (three-dimensional) volume. Due to rotational invariance, the $B$-dependence of $f$ is to leading order quadratic, characterized by the magnetic susceptibility of the QCD vacuum,
\be
\chiM = -\left.\frac{\partial^2 f}{\partial (eB)^2}\right|_{eB=0},
\label{eq:defchitotal}
\ee
which is a dimensionless quantity (here $e>0$ denotes the elementary charge). 
A positive susceptibility indicates a decrease in $f$ due to the magnetic field, that is to say, a {\it paramagnetic} response. On the other hand, $\chiM<0$ is referred to as \emph{diamagnetism}~\cite{springerlink:10.1007/BF01397213}.

In the functional integral formalism of QCD, the susceptibility is readily split into spin- and orbital angular momentum-related terms~\cite{Bali:2012jv}, according to
\be
\chiM=\sum_f\chiM_f,\quad\quad \chiM_f=\chiM^S_f+\chiM^L_f,
\label{eq:chiSO}
\ee
with contributions from each quark flavor $f$ with electric charge $q_f$ and mass $m_f$. For a constant magnetic field $B=F_{xy}$ in the positive $z$ direction,
\be
 \chiM^S_f = \frac{q_f/e}{2m_f}  
 \left.\frac{\partial} {\partial (eB)}\expv{\bar\psi_f\sigma_{xy}\psi_f}\right|_{eB=0},
 \quad\quad\smn = \frac{1}{2i} [\gamma_\mu,\gamma_\nu ].
\label{eq:therelation}
\ee
$\chiM^L_f$ is given by an analogous expression with $\sigma_{xy}$ replaced by 
a generalized angular momentum also present for spinless particles~\cite{Bali:2012jv}. Eq.~(\ref{eq:therelation}) constitutes an important relation which, to our know\-ledge, has not been recognized previously in this context. Its derivation from the quark determinant and the corresponding Dirac operator 
is given in Ref.~\cite{Bali:2012jv}.

Eqs.~(\ref{eq:chiSO}) and~(\ref{eq:therelation}) also show how the magnetic response of the free energy density is related to the breaking of Lorentz symmetry by the external magnetic field. The presence of the preferred direction ($B\parallel z$) induces a nonzero expectation value for the tensor polarization operator $\bar\psi_f\smn\psi_f$, which appears in the spin contribution to the susceptibility, Eq.~(\ref{eq:therelation}). 
To leading order, this expectation value is proportional to the field strength, and, thus, can be written as~\cite{Ioffe:1983ju}
\be
\expv{ \bar\psi_f \sigma_{xy} \psi_f } = q_f B \cdot \expv{\bar \psi_f\psi_f}\cdot \chiF_f \equiv q_f B \cdot \TAU,
\label{eq:defchi}
\ee
where the expectation value of the quark condensate appeared. In the literature, $\chiF_f$ is referred to as the \emph{magnetic susceptibility of the condensate} (for the quark flavor $f$). In what follows, we will also use the term ``magnetic susceptibility''. Again, we stress that it constitutes only one of the two contributions to the total susceptibility. We also define the \emph{tensor coefficient} $\TAU$ as the product of the condensate and the magnetic susceptibility. At finite quark masses, it is advantageous to work with $\TAU$ instead of $\chiF_f$ for reasons related to renormalization (see Sec.~\ref{sec:observables}). 

The magnetic susceptibility $\chiF_f$, in the context of QCD, was first introduced in Ref.~\cite{Ioffe:1983ju}. 
It is a highly relevant quantity for experiments, since it appears in various effects 
\cite{Colangelo:2005hv,Czarnecki:2002nt,Braun:2002en,Pire:2009nn,Nyffeler:2009uw}. In the past, the magnetic susceptibility has been calculated using different methods \cite{Belyaev:1984ic,Balitsky:1985aq,Ball:2002ps,Bergman:2008sg,Gorsky:2009ma,Vainshtein:2002nv,Kim:2004hd,Dorokhov:2005pg,Goeke:2007nc,Nam:2008ff,Ioffe:2009yi,Frasca:2011zn}.
On the lattice, the numerical value of $\chiF_f$ was determined recently in the quenched approximation of two-~\cite{Buividovich:2009ih} and of three-color QCD~\cite{Braguta:2010ej}, in both cases without renormalization. In this talk we report about new results regarding the susceptibility, at physical quark masses in full QCD~\cite{Bali:2012jv}.

An important remark regarding the magnetic susceptibility, which was not mentioned in Ref.~\cite{Bali:2012jv}, is in order here.
At vanishing temperature, electric charge renormalization ensures that the free energy to $\mathcal{O}(B^2)$ is given solely by the energy of the external field itself, $B_r^2/2$, where $B_r$ is the renormalized magnetic field, see e.g.\ \cite{Dunne:2004nc,Endrodi:2013cs}. Since this term is independent of the properties of the medium (i.e., of the QCD vacuum), we do not take it into account, resulting in a vanishing zero-temperature susceptibility,
\be
\chiM(T=0) = 0,
\ee
showing that at $T=0$, the spin- and orbital momentum contributions are equal and of opposite sign. At $T>0$ this relation does not hold anymore, and the two terms become independent. Note that $\mathcal{O}(B^4)$ terms are already present in the free energy at $T=0$, but these do not contribute to $\chiM$. 

\section{Observables and renormalization}
\label{sec:observables}

To realize the external magnetic field on the lattice, we implement the continuum $\mathrm{U}(1)$ gauge field $A_\mu$ satisfying $\partial_x A_y - \partial_y A_x = B$, using space-dependent complex phases (for our implementation, see~\cite{Bali:2011qj}). 
This discretization satisfies periodic boundary conditions in the spatial directions, and ensures that the magnetic flux across the $x-y$ plane is constant (it is also quantized).

We consider three quark flavors $u,d$ and $s$. Since the charges and masses of the quarks differ, we have to treat each flavor separately; $q_u=-2q_d=-2q_s$. Furthermore, we assume $m_u=m_d\ne m_s$. The partition function in the staggered formulation then reads,
\be
\ZZ = \int \D U e^{-\beta S_g} \prod_{f=u,d,s} \left[ \det M(U,q_fB,m_f)\right]^{1/4}, 
\label{eq:partfunc}
\ee
with $M(U,qB,m) = \slashed D(U,qB) + m\mathds{1}$ being the fermion matrix and $\beta=6/g^2$ the gauge coupling. The exact form of the action we use and details of the simulation setup are given in Ref.~\cite{Bali:2011qj} and in references there.

In this formulation, the expectation value of the quark condensate for the flavor $f$ can be written as
\be
\expv{\bar{\psi}_f\psi_f} \equiv \frac{T}{V}\frac{\partial \log \ZZ}{\partial m_f} = \frac{T}{4V} \expv{\Tr\, M^{-1}(U,q_fB,m_f)}.
\label{eq:pbpdef}
\ee
Likewise, the expectation value of the tensor Dirac structure of Eq.~(\ref{eq:defchi}) reads,
\be
\expv{\bar\psi_f \smn \psi_f} = \frac{T}{4V} \expv{\Tr \, M^{-1}(U,q_fB,m_f)\smn}.
\label{eq:pbpTdef}
\ee

In order to determine the continuum limit of the above observables, their renormalization has to be performed.
The quark condensate (at finite mass) is subject to additive and multiplicative renormalization. All of these divergences -- being independent of the temperature and of the magnetic field -- cancel~\cite{Bali:2011qj} in the following combination~\cite{Bali:2012zg},
\be
\Sigma_{u,d}(B,T) = \frac{2m_{ud}}{M_\pi^2 F^2} \left[ \bar\psi\psi_{u,d}(B,T)- \bar\psi\psi_{u,d}(0,0) \right] + 1, 
\label{eq:pbpren}
\ee
where, to obtain a dimensionless quantity, we divided by the combination $M_\pi^2 F^2$ which contains the zero-field pion mass $M_\pi=135 \textmd{ MeV}$ and (the chiral limit of the) pion decay constant $F=86 \textmd{ MeV}$~\cite{Colangelo:2003hf}. This specific combination enters the Gell-Mann-Oakes-Renner relation,
\be
2m_{ud} \cdot \bar\psi\psi(0,0) = M_\pi^2 F^2+\cdots.
\ee
The factor $+1$ is included in Eq.~(\ref{eq:pbpren}), so that the chiral limit of the condensate is fixed to 1 at $T=B=0$, and approaches 0 as $T\to\infty$. At nonzero quark mass, $\Sigma_{u,d}$ will still start from 1 at $T=B=0$. 

We also define the change of the renormalized condensate due to the magnetic field as
\be
\Delta\Sigma_{u,d}(B,T) = \Sigma_{u,d}(B,T)-\Sigma_{u,d}(0,T)
 = \frac{2m_{ud}}{M_\pi^2 F^2} \left[ \bar\psi\psi_{u,d}(B,T)- \bar\psi\psi_{u,d}(0,T) \right].
\label{eq:deltapbp}
\ee
Note that the $\bar\psi\psi(0,0)$ term has canceled from this difference.
Again, our normalization is such that the change of the condensate is measured in units of the chiral condensate at $B=0$ and $T=0$. This normalization will be advantageous when comparing the lattice results to $\chi PT$ and model predictions, which are usually given in units of $\bar\psi\psi(0,0)$. 

The renormalization properties of the tensor polarization are somewhat different.
As a calculation in the free theory shows, an additive divergence of the form $\zetaT \cdot q_fB \cdot m_f \log(m_f^2a^2)$ appears in $\expv{\bar\psi_f\smn\psi_f}$~\cite{Bali:2012jv}. Moreover, there is a multiplicative divergence as well, which will be canceled by the tensor renormalization constant $\ZT$. Both $\zetaT$ and $\ZT$ are independent of $T$ and $B$ (and in mass-independent schemes of $m_f$). In the free theory, the coefficient of the logarithmic divergence is calculated to be $\zetaT(g=0)=3/(4\pi^2)$~\cite{Bali:2012jv}.

Due to the divergence structure of the condensate and of the tensor polarization, it is advantageous to consider the tensor coefficient $\TAU$ defined in Eq.~(\ref{eq:defchi}). We notice that the operator $1-m_f\partial/\partial m_f$ eliminates the logarithmic divergence, and, thus, can be used to define an observable with a finite continuum limit,
\be
\TAU^r \equiv \left(1-m_f\frac{\partial}{\partial m_f}\right)\TAU \cdot \ZT \equiv \TAU \ZT - \TAU^{\rm div}.
\label{eq:mdmsub}
\ee
At finite quark mass, this is one possible prescription to cancel the additive logarithmic term. 
It has the advantages that the chiral limit of $\TAU$ is left unaffected, and that, together with the logarithmic divergence, scheme-dependent finite terms also cancel in this difference~\cite{Bali:2012jv}, such that the scheme- and renormalization scale-dependence of $\TAU$ resides solely in $\ZT$. The latter renormalization constant we calculate perturbatively, in the $\MSBar$ scheme at a renormalization scale of $\mu=2\textmd{ GeV}$. For the details of the perturbative calculation, see Ref.~\cite{Bali:2012jv}.

\section{Results}

\begin{figure}[t]
\centering
\includegraphics*[width=8cm]{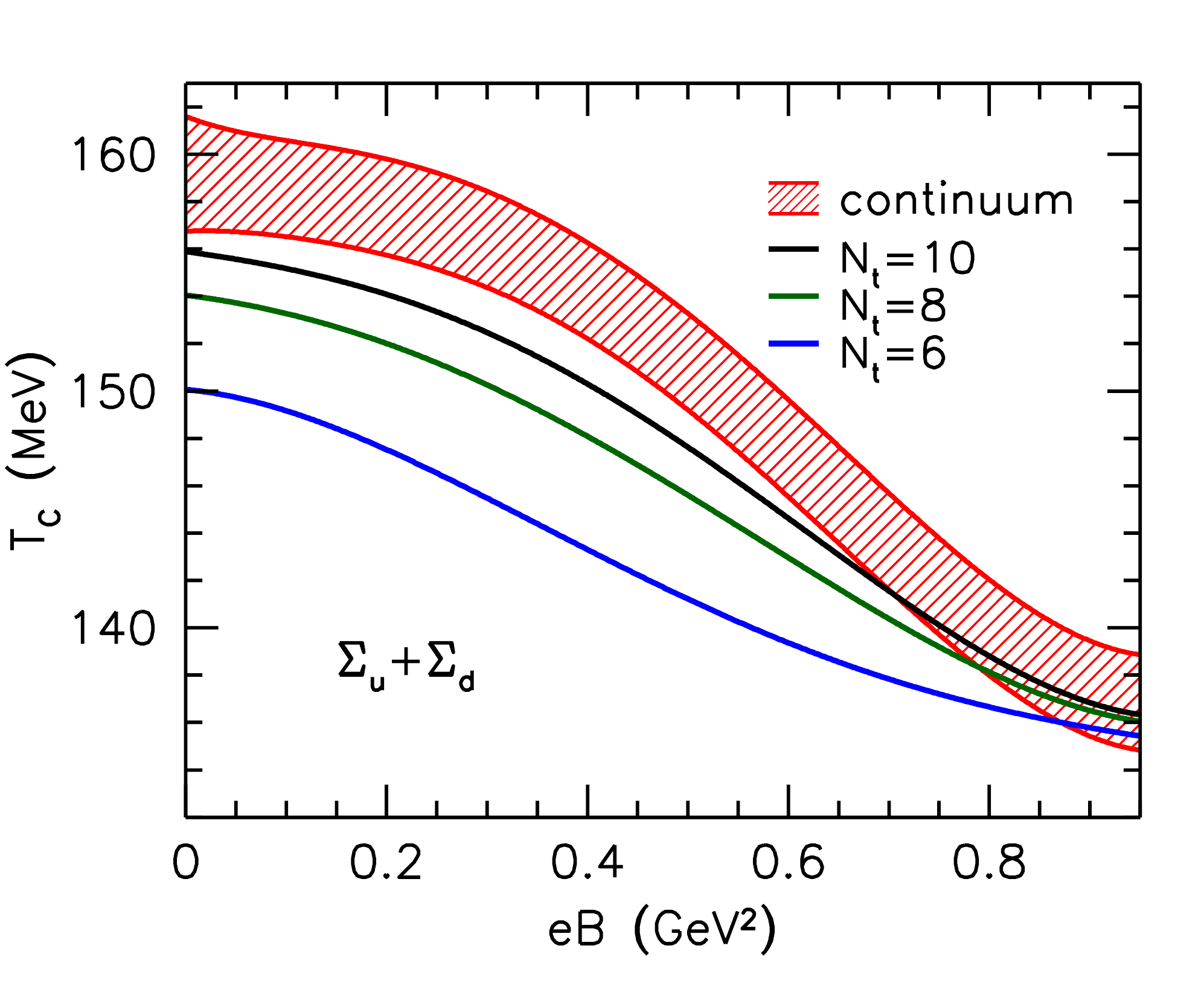}
\caption{QCD phase diagram in the magnetic field-temperature plane. Shown is the transition temperature $T_c(B)$ obtained from the renormalized light condensates, Eq.~(\protect\ref{eq:pbpren}), averaged over the $u$ and $d$ quark.}
\label{fig:phase_diagram}
\end{figure}

We start with our main result, the QCD phase diagram in the $B-T$ plane, Fig.\ \ref{fig:phase_diagram}. The transition temperature from the renormalized condensate of the light quarks  {\it decreases} by up to $20$~MeV for magnetic fields up to $eB\simeq 1$ GeV$^2$. A similar behavior is found in the chiral susceptibility and the strange quark number susceptibility as well \cite{Bali:2011qj}. We stress that these quantities are given in the continuum limit extrapolated from $N_t=6,\,8,\,10$ lattices using physical quark masses.

\begin{figure}[!b]
\centering
\mbox{
\includegraphics*[width=6.0cm]{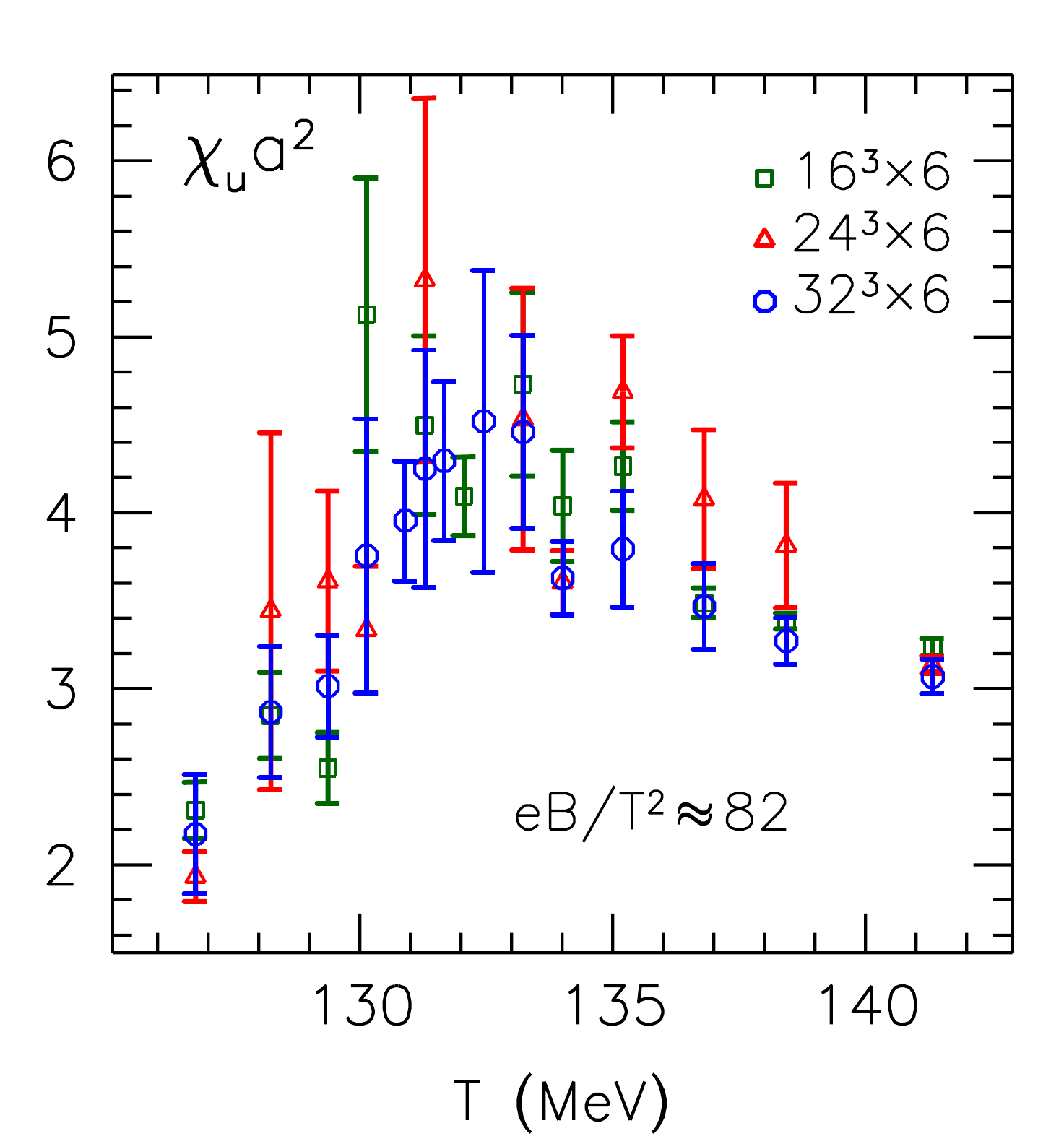}\qquad
\includegraphics*[width=8.0cm]{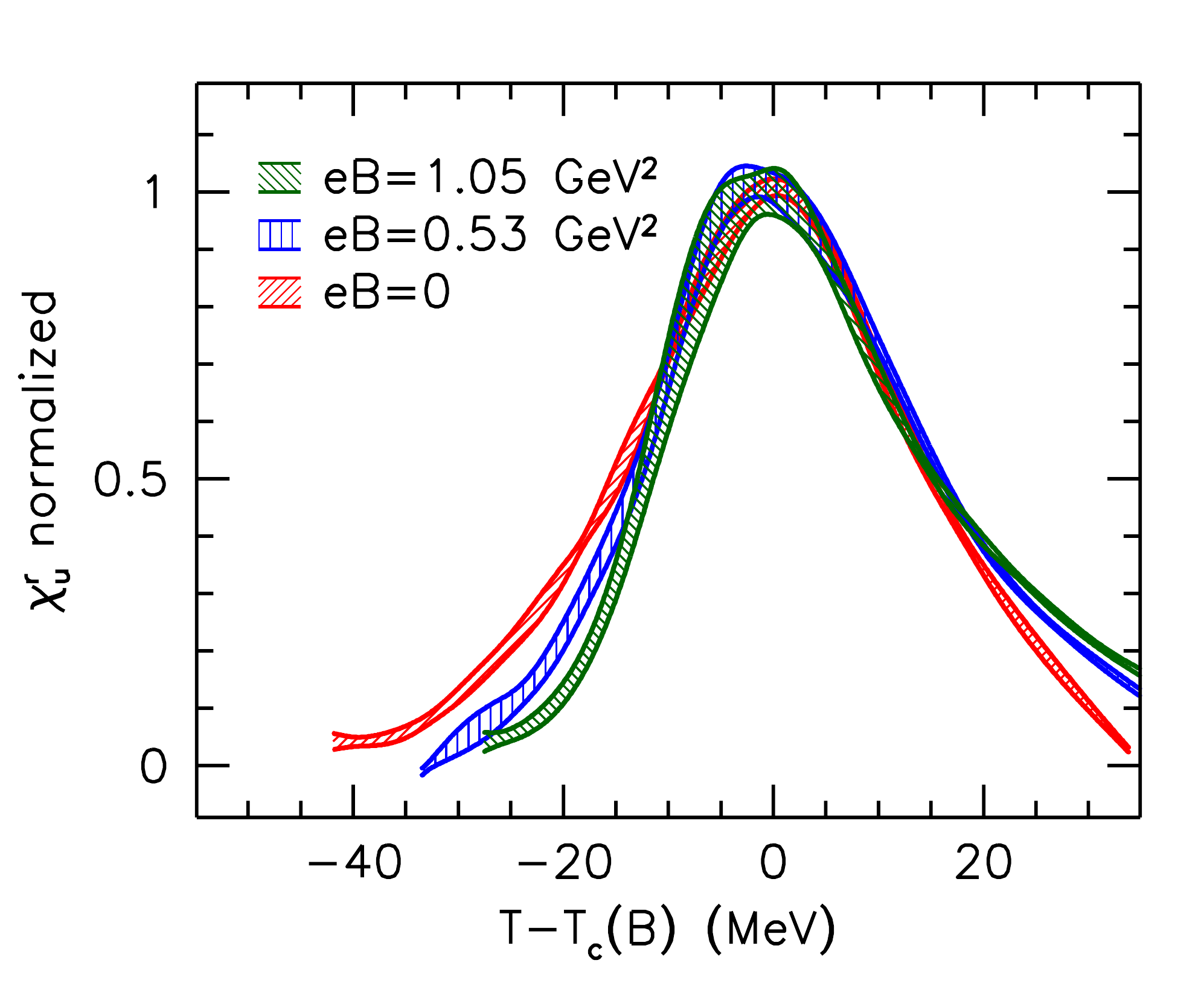}}
\caption{Strength of the transition analyzed through the light susceptibility $\chi_u$ on $N_t=6$ lattices, for details see \cite{Bali:2011qj}. Neither the volume scaling (left) nor the relative change in the peak shape (right) show evidence for a strengthening of the transition with the magnetic field, up to $eB=1$ GeV$^2$.}
\label{fig:nature}
\end{figure}

The transition remains a crossover up to our largest magnetic fields as we verify by the finite volume scaling and the relative change in the $T$-dependence of the chiral susceptibility, Fig.~\ref{fig:nature}.

\begin{figure}[t]
\centering
\mbox{
\includegraphics*[width=7.7cm]{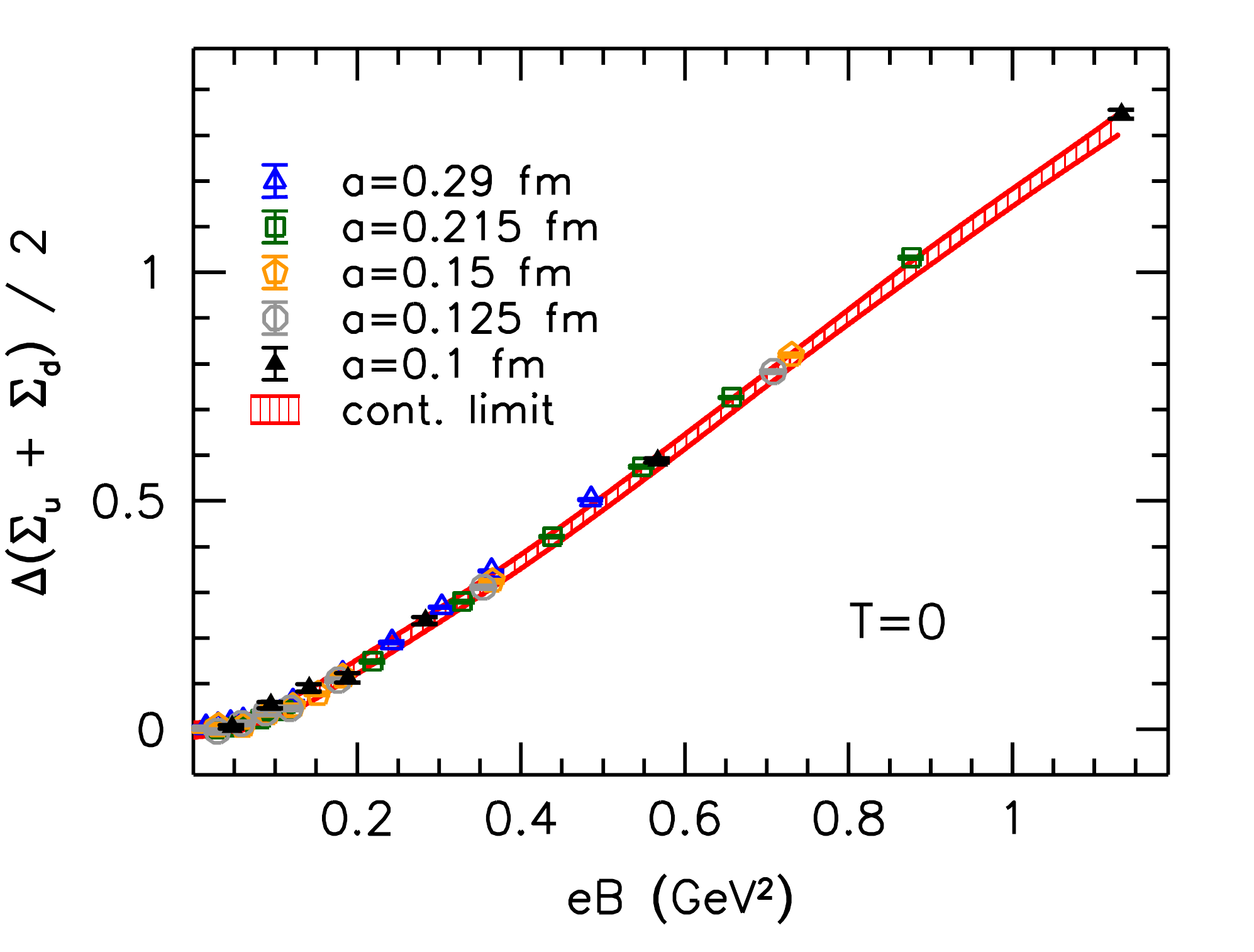}\quad
\includegraphics*[width=7.7cm]{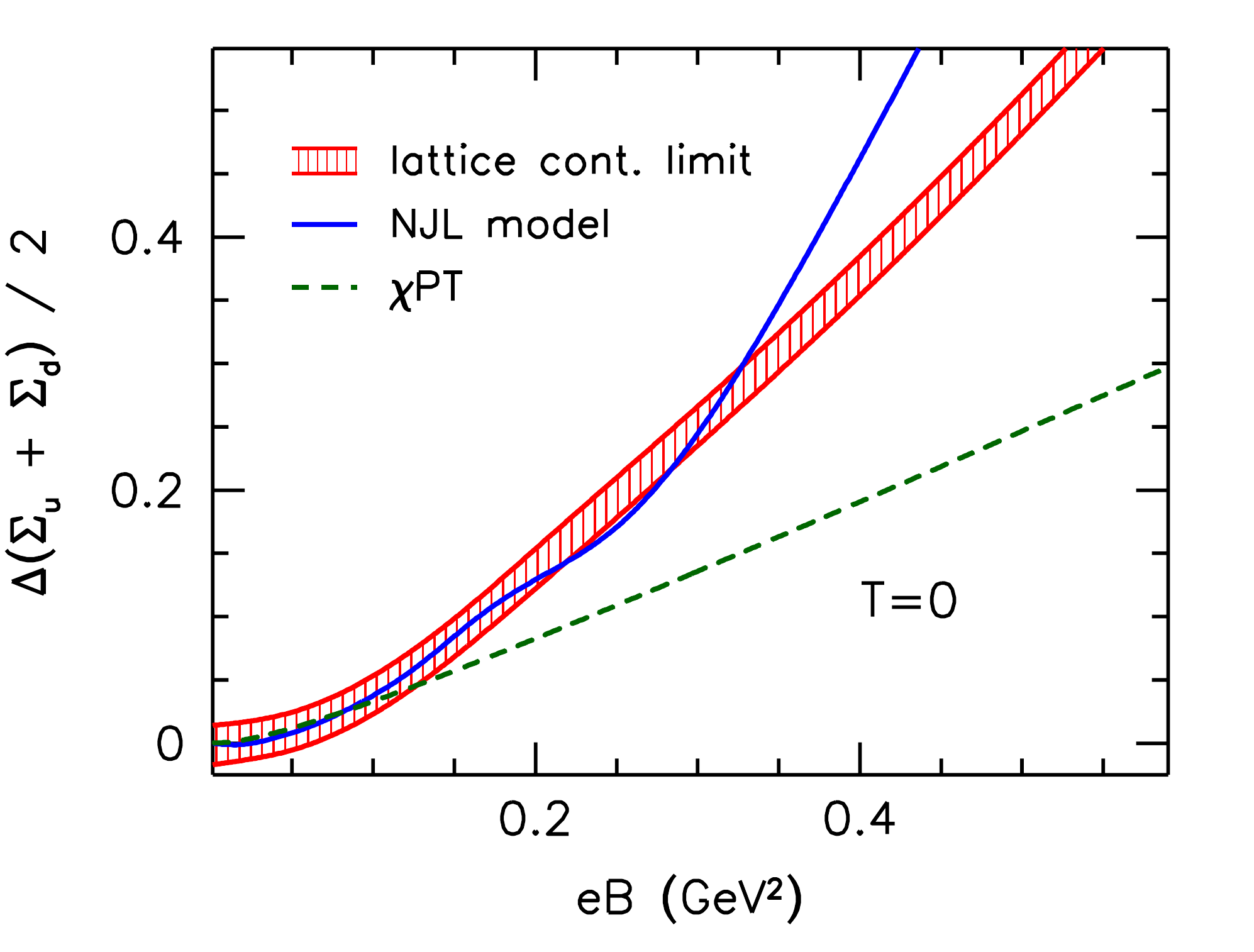}}
\caption{Magnetic catalysis at zero temperature. Shown is the change in the average of the light condensates, Eq.~(\protect\ref{eq:deltapbp}), due to the magnetic field using five lattice spacings and the extrapolation to the continuum limit (left) and then in comparison with the NJL model \cite{Gatto:2010pt} and chiral perturbation theory \cite{Cohen:2007bt,Andersen:2012dz} in the low-$B$ region (right).}
\label{fig:MC}
\end{figure}

We start our discussion of the light condensate at
zero temperature. The two panels of Fig.~\ref{fig:MC} show our lattice data and the extrapolation to the continuum limit, and a comparison to the NJL model and chiral perturbation theory. We confirm the increase of $\expv{\bar \psi_f\psi_f}$ with growing magnetic field, the well-known magnetic catalysis. The NJL and $\chi$PT predictions agree quantitatively with our data for $eB \lesssim 0.3$~GeV$^2$ and $eB \lesssim 0.1$~GeV$^2$, respectively. The limitation of these approaches is not unexpected. The breakdown of the  
$\chi$PT prediction, for example, may be understood from the fact that the QCD dynamics is less and less dominated by pions as $B$ grows, 
and charged $\rho^\pm$ mesons play a non-negligible role, cf.~\cite{Endrodi:2013cs,Chernodub:2010qx}.

\enlargethispage{\baselineskip}
\begin{figure}[b!]
\centering
\mbox{\includegraphics*[width=7.7cm]{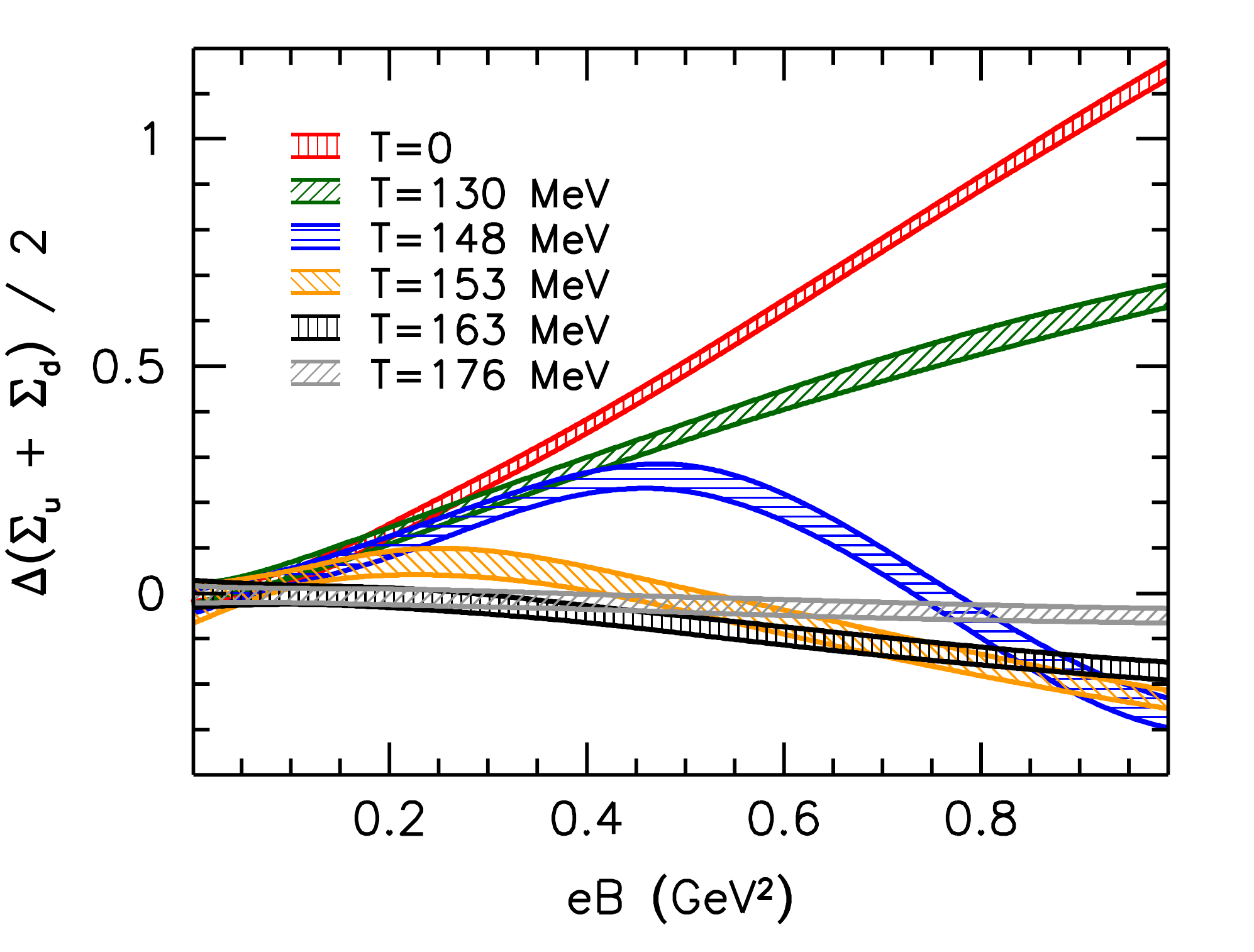}\quad
\includegraphics*[width=7.7cm]{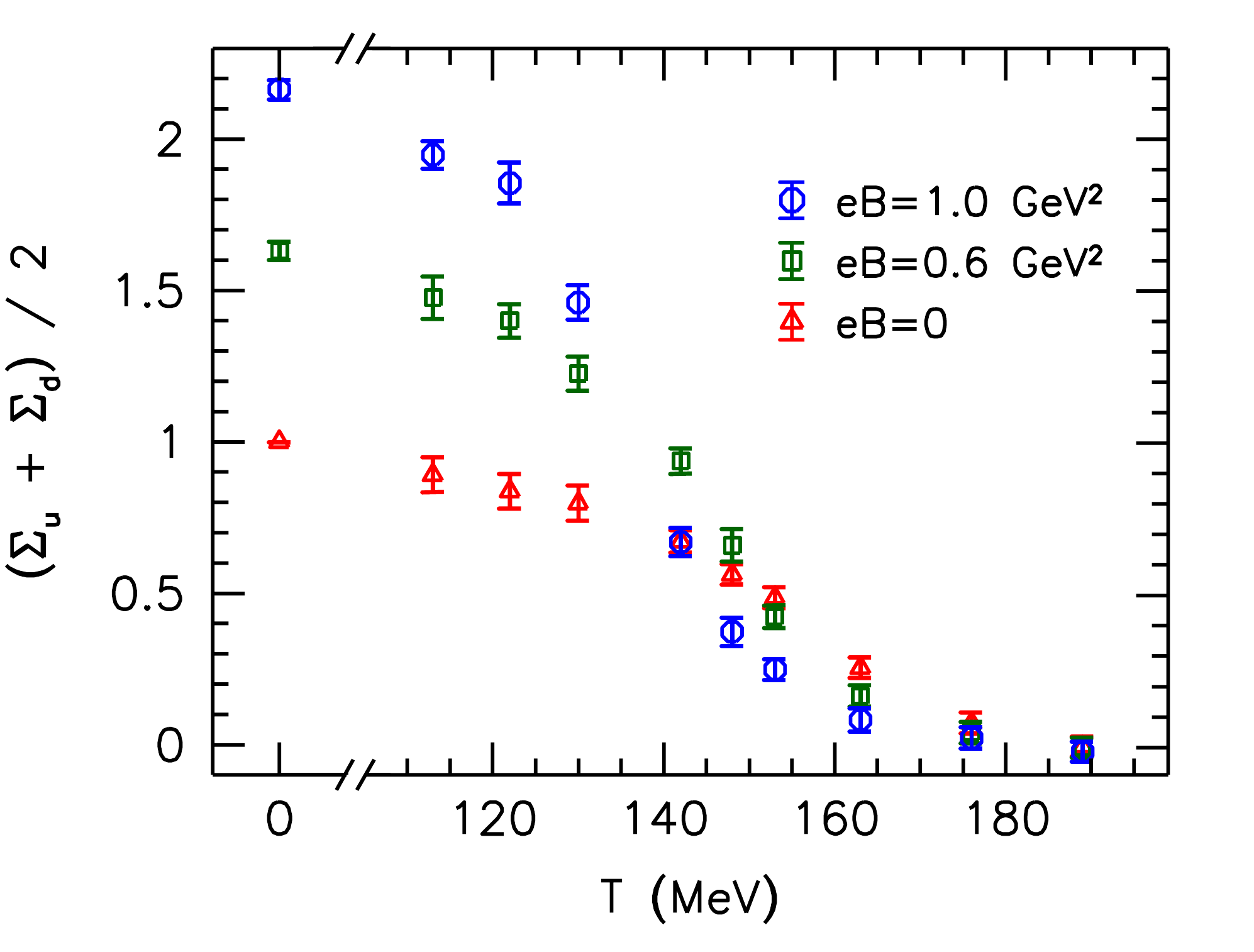}}
\caption{Inverse magnetic catalysis at the QCD transition. The light condensate as a function of the magnetic field for fixed temperatures (left) and as a function of $T$ for fixed $B$ (right) displays a non-monotonic behavior with MC at zero temperature turning into IMC around $T_c$. The transition temperatures $T_c(B)$ of this observable are obtained from the inflection points in the right panel and hence can be seen to decrease with~$B$. The right plot is the continuum extrapolation of Fig.~1 in \protect\cite{Bali:2011uf} (where a slighty different normalization was used).}
\label{fig:IMC}
\end{figure} 

For temperatures around the transition, we show the dependence of the light condensate on the magnetic field in Fig.~\ref{fig:IMC}. As advertised, the behavior is non-monotonic. In the temperature range 150-175 MeV the condensate decreases with the magnetic field, which is the inverse magnetic catalysis. It is especially this behavior that is missing in almost all effective theories of QCD, see e.g.\ the finite temperature comparison to a PNJL model and $\chi$PT in Fig.~4 of \cite{Bali:2012zg}; for a recent attempt see \cite{Fukushima:2012kc}. From the right panel of Fig.~\ref{fig:IMC}, the decrease of the transition temperature with the magnetic field is visible. 

\begin{figure}[t]
\centering
\mbox{
\includegraphics[width=0.5\linewidth]{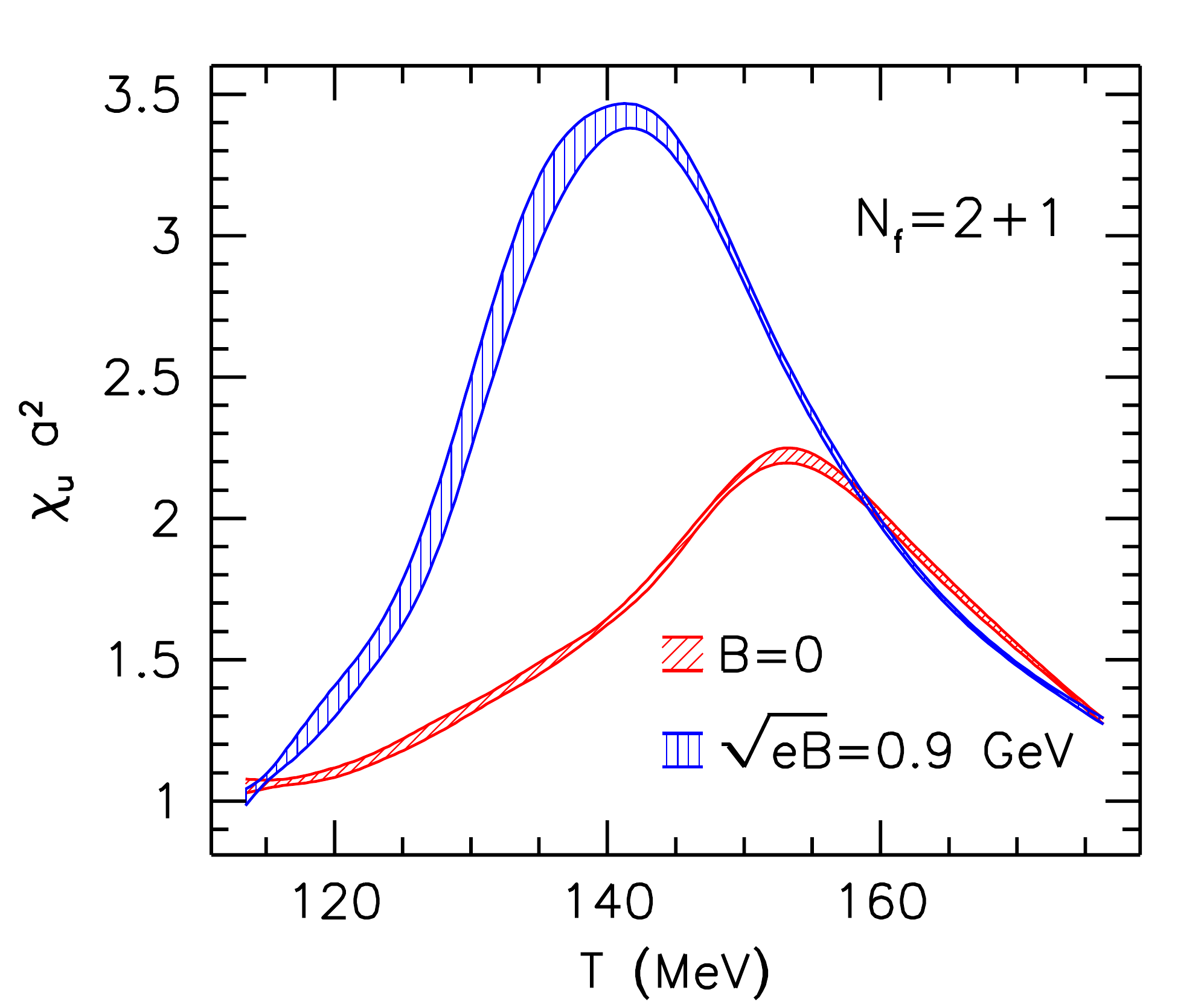}\quad
\includegraphics[width=0.5\linewidth]{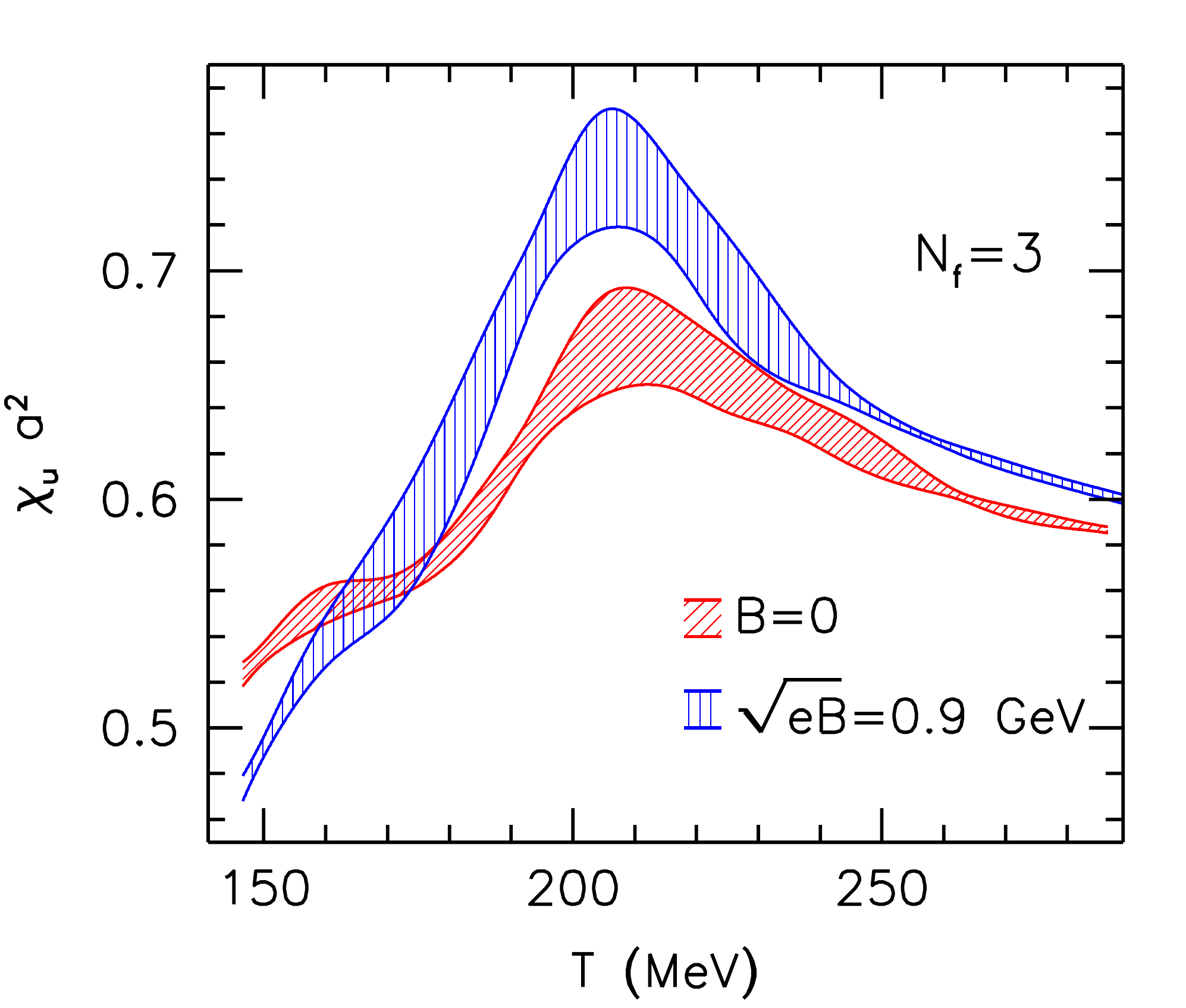}}
\caption{Influence of the quark mass on the dependence $T_c(B)$ of the (bare) light susceptibility on $N_t=6$ lattices. For physical masses (left) the maximum moves to lower temperatures with the magnetic field, whereas in the system with three quarks at the strange quark mass (right) no change in position of the maximum is visible. This means that the effect of a decrease of $T_c$ with $B$ is washed out by heavier-than-physical quark masses.}
\label{fig:mass_dep}
\end{figure}

For the comparison to other lattice simulations of finite temperature gauge theories with magnetic fields in $SU(3)$ \cite{D'Elia:2010nq} and $SU(2)$ \cite{Ilgenfritz:2012fw}, we first stress that our lattice results have been extrapolated to the continuum limit. Another crucial difference to those simulations is the mass of the quarks, which is physical only in our simulations. The consequences of this we have investigated by artificially lifting the masses of the light quarks to the strange quark mass in our system, performing new simulations. As Fig.~\ref{fig:mass_dep} shows by virtue of the light susceptibility, no decrease in $T_c$ with the magnetic field can be found anymore. The effect of IMC is washed out by the heavy quarks as well, cf.\ Fig.~5 lower panels  in \cite{Bali:2011qj}. This mass dependence should, at least partially, be responsible for the different results obtained in other lattice simulations \cite{D'Elia:2010nq,Ilgenfritz:2012fw}. 

\begin{figure}[t!]
\centering
\mbox{
\includegraphics*[width=7.7cm]{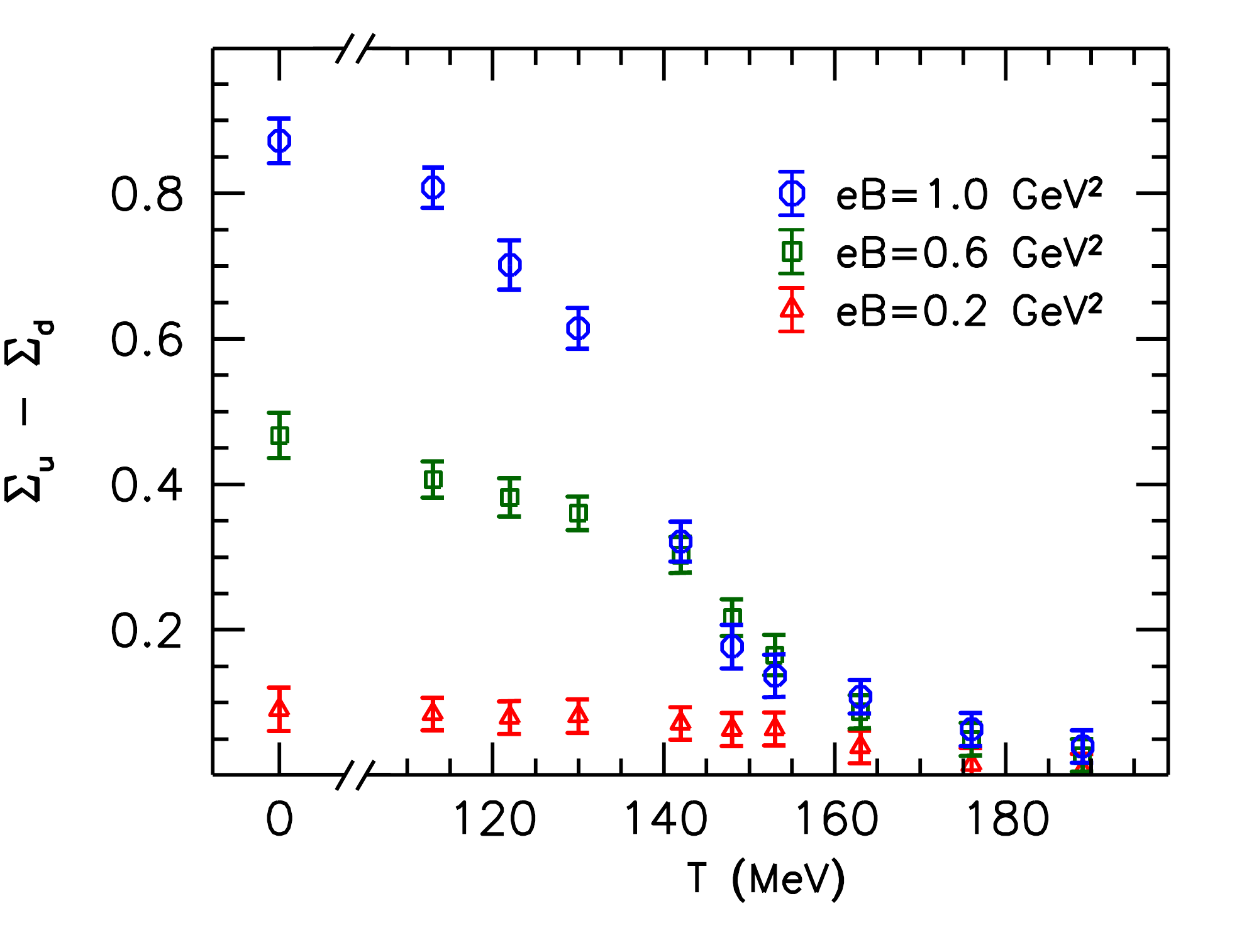}}
\caption{Difference of the light condensates, $\Delta\Sigma_u-\Delta\Sigma_d = \Sigma_u-\Sigma_d$ signalling isospin breaking, as a function of temperatures for different magnetic fields, in analogy to Fig.~\protect\ref{fig:IMC} right panel.}
\label{fig:isospin}
\end{figure}

With the mass-degenerate light quarks employed here, isospin symmetry is exact at vanishing field. At non-vanishing magnetic fields, however, this symmetry is broken by the quark charges $-2e/3$ vs. $e/3$. Therefore, we also consider the difference of the light condensates, $\Delta\Sigma_u-\Delta\Sigma_d = \Sigma_u-\Sigma_d$. As Fig.~\ref{fig:isospin} shows, this observable takes on nonzero expectation values and its temperature dependence is similar to that of the average light condensate.

\bigskip
\enlargethispage{-2\baselineskip}

In the second part of this section, we present lattice data on the tensor polarization, again continuum extrapolated, and compute the magnetic susceptibility $\chiF_f$ and the tensor coefficient $\TAU$ as the leading order proportionality factors in Eq.~(\ref{eq:defchi}).

\begin{figure}[b]
\centering
\mbox{
\includegraphics*[width=7.7cm]{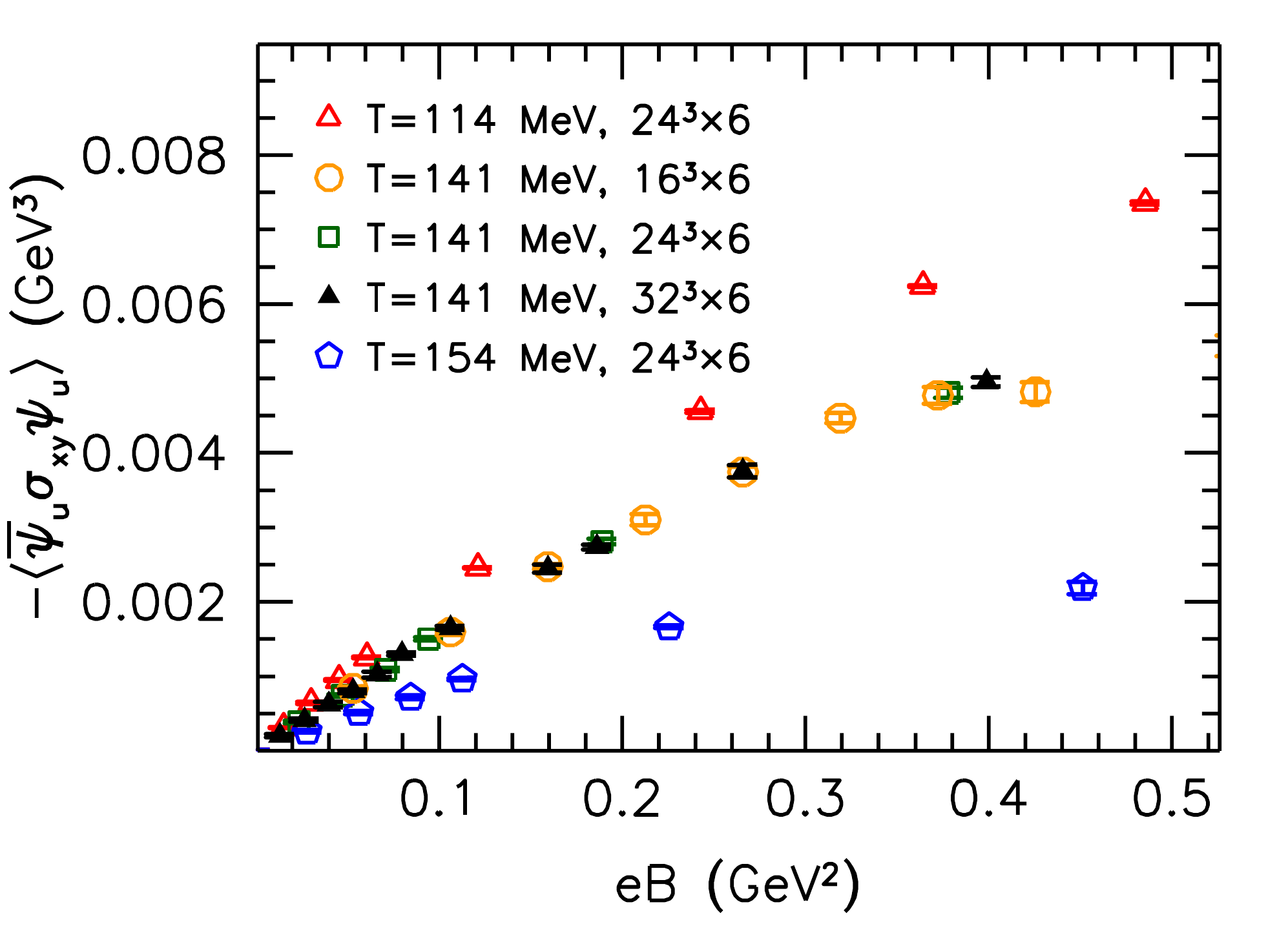}}
\caption{Negative tensor polarization as a function of the magnetic field for three temperatures near the transition. The agreement between the data for different spatial volumes at $T=141$ MeV indicates negligible finite volume effects.}
\label{fig:tensor_pre}
\end{figure}

First of all, Fig.~\ref{fig:tensor_pre} clearly reveals a linear growth of the tensor polarization 
$\expv{ \bar\psi_f \sigma_{xy} \psi_f }$ with the magnetic field. Already from this figure it is clear that both $\chiF_f$ and $\TAU$ are negative, and that the spin contribution to the QCD vacuum is, thus, of {\it diamagnetic} nature. This holds for all temperatures up to 200 MeV. In solid state physics, where one is interested in the nature of positive energy electrons, the spin is associated with paramagnetism. Here, however, we analyze the response of the QCD vacuum, which is different, because, e.g., the vacuum energy for fermions is negative. For reviews about the subject we refer the reader to Refs.~\cite{Nielsen:1980sx,Wilczek:1996bw,Grozin:2008yd}.

\begin{figure}[t]
\centering
\mbox{
\includegraphics*[width=7.7cm]{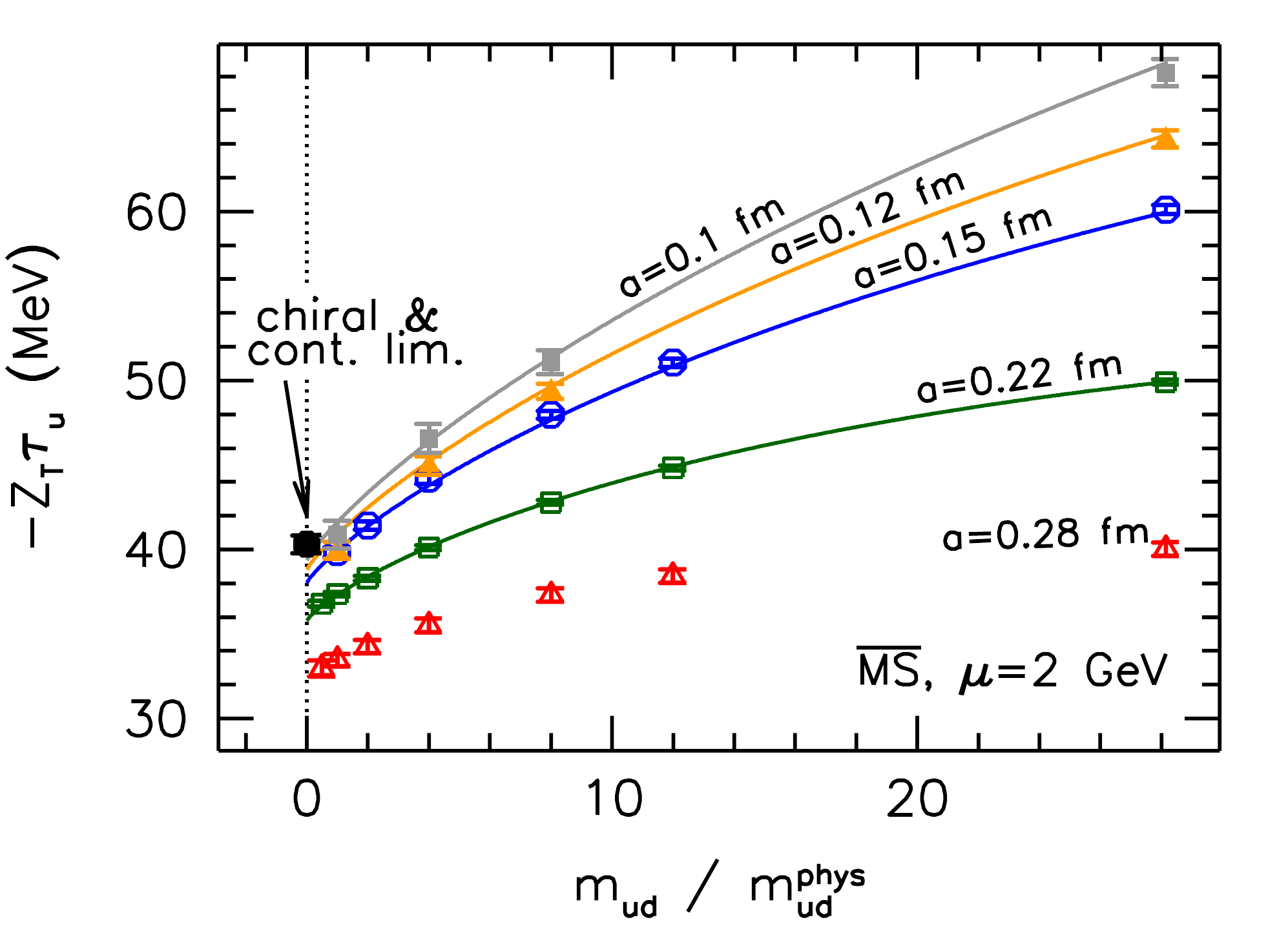}
}
\caption{Mass dependence of the combination $-\ZT\tau_u$ in the $\MSBar$ scheme at renormalization scale $\mu=2$ GeV. The coefficient of the logarithmic divergence is determined by fitting the data by Eq.~(\protect\ref{eq:tensor_fit}) (solid lines).}
\label{fig:tensor}
\end{figure}

The corresponding tensor coefficients at zero temperature are plotted in Fig.~\ref{fig:tensor} as a function of the light quark masses, that we varied between the physical value and the strange quark mass, again through new dynamical simulations. These data have already been multiplied by the tensor renormalization constant $\ZT$ (and $-1$), but still contain a divergence of the form $\log(m_f a)$ as we have discussed before Eq.~(\ref{eq:mdmsub}). Correspondingly, we fit these data with coefficients $c$ using 
\be
 c_0 +c_1 R +c_2 R \log(R^2 a^2)\,,\quad 
 R = m_{ud}/ m_{ud}^{\rm phys}\,,\quad 
 c_i = c^{(0)}_i + c^{(1)}_i a^2\,,
\label{eq:tensor_fit}
\ee
such that 
\be
 \TAU^r = c_{f0} + 2 c_{f2} m_f\,.
\ee
We remark that the coefficients of the logarithms, $c^{(0)}_{u2}/m_{ud}^{\rm phys}=0.055(5)$ and $c^{(0)}_{d2}/m_{ud}^{\rm phys}=0.072(6)$ are quite close to the free-field value of $3/(4\pi^2)$. As a result of these fits, we get the following values for the renormalized tensor coefficients in the chiral limit,
\be
\tau^r_u = -40.3(1.4)\, {\rm MeV}\,,\qquad
\tau^r_d = -38.9(1.5)\, {\rm MeV}\,,
\ee
and at physical quark masses,
\be
\tau^r_u = -40.7(1.3)\, {\rm MeV}\,,\qquad
\tau^r_d = -39.4(1.4)\, {\rm MeV}\,,\qquad
\tau^r_s = -53.0(7.2)\, {\rm MeV}\,.
\ee

In order to extract the magnetic susceptibilities in the $\MSBar$ scheme at $\mu=2$ GeV, we use a recent lattice determination of $\expv{\bar{\psi}_l\psi_l}=(269(2)\,{\rm MeV})^3$ \cite{Durr:2010vn} and $\expv{\bar{\psi}_s\psi_s}/\expv{\bar{\psi}_l\psi_l}=0.8(3)$ from \cite{Jamin:2002ev}. This yields 
\be
 \chi_u = - (2.08\pm 0.08)\, {\rm GeV}^{-2}\,,\qquad
 \chi_d = - (2.02\pm 0.09)\, {\rm GeV}^{-2}\,,\qquad
 \chi_s = - (3.4\pm 1.4)\, {\rm GeV}^{-2}\,.
\ee

These values are in good agreement with the QCD sum rule calculations summarized and updated in \cite{Ball:2002ps}, $\chi_{ud}=-2.11(23)\, {\rm GeV}^{-2}$. Quenched lattice simulations, on the other hand, gave the unrenormalized values
$\chi_{ud}=-1.547(6)\, {\rm GeV}^{-2}$ for two colors \cite{Buividovich:2009ih} and $\chi_{ud}=-4.24(18)\, {\rm GeV}^{-2}$ for three colors \cite{Braguta:2010ej}.

\begin{figure}[t!]
\centering
\mbox{
\includegraphics*[width=7.7cm]{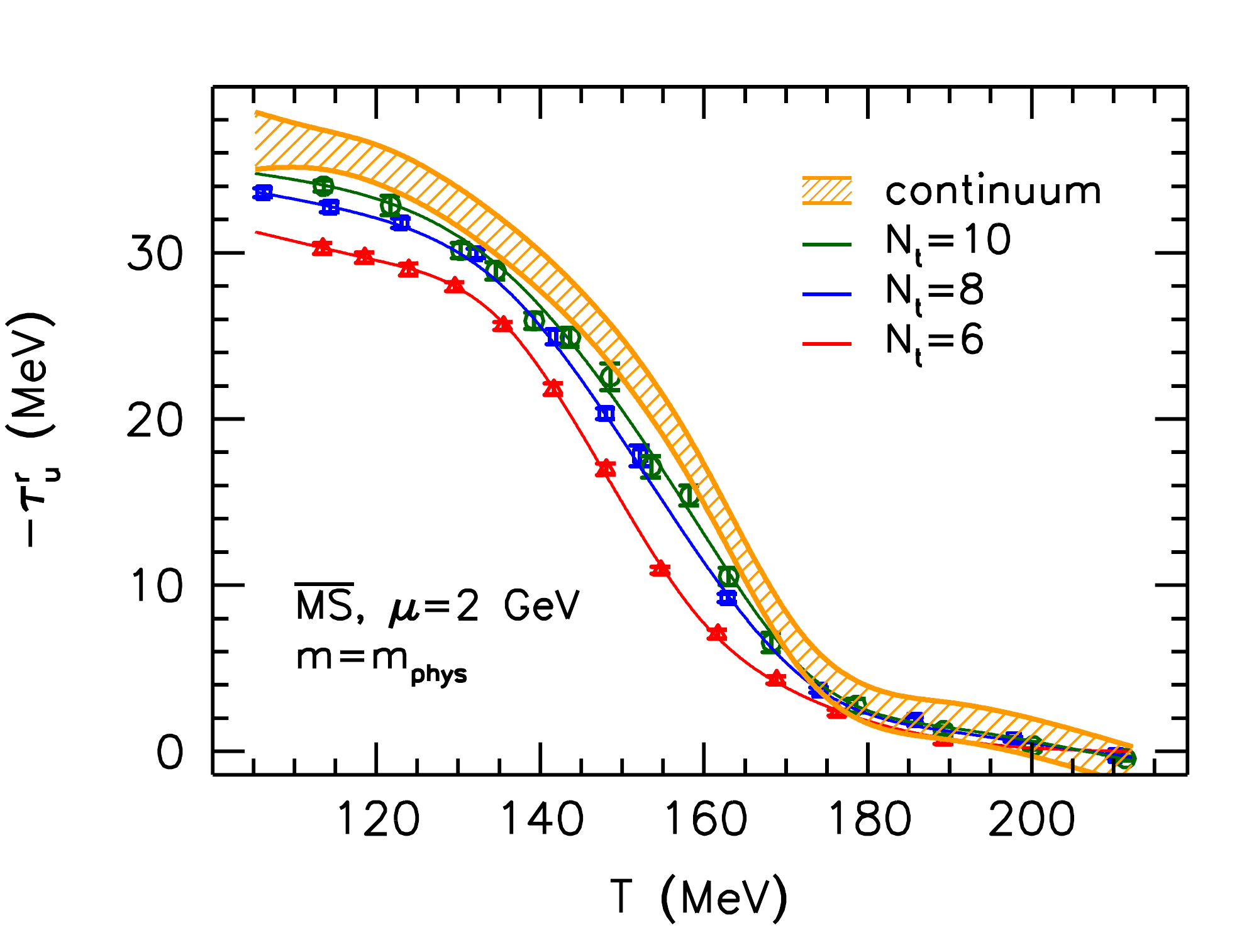}
}
\caption{Temperature dependence of the negative renormalized light tensor coefficient in the $\MSBar$ scheme at a renormalization scale
$\mu=2$ GeV for three lattice spacings and the continuum
extrapolation.}
\label{fig:tensor_temp}
\end{figure}

Finally, we show the corresponding temperature dependence of $\tau^r_u$ in Fig.~\ref{fig:tensor_temp}. 
Since the divergent term $\tau^{\rm div}$ of Eq.~(\ref{eq:mdmsub}) is independent of $T$, we subtracted it in the same way as at $T=0$. Then we
performed a combined fit of the form $N_t^{-2}$ for different lattice spacings. We found the light tensor coefficient to decrease in the transition region like a quasi-order parameter. The corresponding transition temperature, 
$T_c=162(3)(3)$ MeV, is similar to that of the chiral condensate.

\section{Summary}

We have presented results on equilibrium thermodynamics of QCD in external magnetic fields, obtained through lattice simulations. We have found the well-known magnetic catalysis at zero temperature to turn into the opposite -- inverse magnetic catalysis -- at temperatures around the transition. We have argued that performing the continuum limit and using physical quark masses are essential for obtaining these results.  The inverse magnetic catalysis behavior still awaits confirmation from other first principle methods, for instance from lattice simulations with other fermion discretizations. This also applies to the associated decrease of the transition temperatures with the magnetic field, which could be of phenomenological relavance for, e.g.,  heavy-ion collisions. Both phenomena seem to represent a challenge for effective theories of QCD. 

As a second key aspect we have calculated one of the Lorentz symmetry breaking observables induced by the magnetic field, the tensor polarization. It gives a diamagnetic contribution to the spin-related part of the magnetic susceptibility of the QCD vacuum. 

\section{Acknowledgements}
Our work was supported by DFG grants BR 2872/4-2, FO 502/1-2 and SFB-TR 55, the EU grants No. PITN-GA-2009-238353 (ITN
STRONGnet) and ERC Grant 208740, and the Cyprus Research Promotion Foundation under Contract No. TECHNOLOGY/$\Theta E\Pi I\Sigma$/0311(BE)/16.

\bibliographystyle{JHEP}
\bibliography{thermomag}

\providecommand{\href}[2]{#2}\begingroup\raggedright\begin{thebibliography}{10}

\bibitem{Aoki:2006we}
Y.~Aoki, G.~Endr{\H{o}}di, Z.~Fodor, S.~D. Katz, and K.~K. Szab{\'o} {\em
  Nature} {\bf 443} (2006) 675,
  [\href{http://xxx.lanl.gov/abs/hep-lat/0611014}{{\tt hep-lat/0611014}}].

\bibitem{Bali:2011qj}
G.~Bali, F.~Bruckmann, G.~Endr\H{o}di, Z.~Fodor, S.~Katz, {\em et.~al.} {\em
  JHEP} {\bf 1202} (2012) 044, [\href{http://xxx.lanl.gov/abs/1111.4956}{{\tt
  arXiv:1111.4956}}].

\bibitem{Bali:2012zg}
G.~Bali, F.~Bruckmann, G.~Endr\H{o}di, Z.~Fodor, S.~Katz, {\em et.~al.} {\em
  Phys. Rev.} {\bf D86} (2012) 071502,
  [\href{http://xxx.lanl.gov/abs/1206.4205}{{\tt arXiv:1206.4205}}].

\bibitem{Bali:2012jv}
G.~Bali, F.~Bruckmann, M.~Constantinou, M.~Costa, G.~Endr\H{o}di, {\em et.~al.}
  {\em Phys. Rev.} {\bf D86} (2012) 094512,
  [\href{http://xxx.lanl.gov/abs/1209.6015}{{\tt arXiv:1209.6015}}].

\bibitem{Schramm:1991ex}
S.~Schramm, B.~Muller, and A.~J. Schramm {\em Mod. Phys. Lett.} {\bf A7} (1992)
  973--982.

\bibitem{Gusynin:1995nb}
V.~P. Gusynin, V.~A. Miransky, and I.~A. Shovkovy {\em Nucl. Phys.} {\bf B462}
  (1996) 249, [\href{http://xxx.lanl.gov/abs/hep-ph/9509320}{{\tt
  hep-ph/9509320}}].

\bibitem{Shovkovy:2012zn}
I.~A. Shovkovy \href{http://xxx.lanl.gov/abs/1207.5081}{{\tt arXiv:1207.5081}}.

\bibitem{Buividovich:2008wf}
P.~Buividovich, M.~Chernodub, E.~Luschevskaya, and M.~Polikarpov {\em Phys.
  Lett.} {\bf B682} (2010) 484, [\href{http://xxx.lanl.gov/abs/0812.1740}{{\tt
  arXiv:0812.1740}}].

\bibitem{Braguta:2010ej}
V.~Braguta, P.~Buividovich, T.~Kalaydzhyan, S.~Kuznetsov, and M.~Polikarpov
  {\em PoS} {\bf LATTICE2010} (2010) 190,
  [\href{http://xxx.lanl.gov/abs/1011.3795}{{\tt arXiv:1011.3795}}].

\bibitem{D'Elia:2010nq}
M.~D'Elia, S.~Mukherjee, and F.~Sanfilippo {\em Phys. Rev.} {\bf D82} (2010)
  051501, [\href{http://xxx.lanl.gov/abs/1005.5365}{{\tt arXiv:1005.5365}}].

\bibitem{D'Elia:2011zu}
M.~D'Elia and F.~Negro {\em Phys. Rev.} {\bf D83} (2011) 114028,
  [\href{http://xxx.lanl.gov/abs/1103.2080}{{\tt arXiv:1103.2080}}].

\bibitem{Ilgenfritz:2012fw}
E.-M. Ilgenfritz, M.~Kalinowski, M.~Muller-Preussker, B.~Petersson, and
  A.~Schreiber {\em Phys. Rev.} {\bf D85} (2012) 114504,
  [\href{http://xxx.lanl.gov/abs/1203.3360}{{\tt arXiv:1203.3360}}].

\bibitem{springerlink:10.1007/BF01397213}
L.~Landau {\em Zeitschrift f\"ur Physik A, Hadrons and Nuclei} {\bf 64} (1930)
  629. 10.1007/BF01397213.

\bibitem{Ioffe:1983ju}
B.~L. Ioffe and A.~V. Smilga {\em Nucl. Phys.} {\bf B232} (1984) 109.

\bibitem{Colangelo:2005hv}
P.~Colangelo, F.~De~Fazio, and A.~Ozpineci {\em Phys. Rev.} {\bf D72} (2005)
  074004, [\href{http://xxx.lanl.gov/abs/hep-ph/0505195}{{\tt
  hep-ph/0505195}}].

\bibitem{Czarnecki:2002nt}
A.~Czarnecki, W.~J. Marciano, and A.~Vainshtein {\em Phys. Rev.} {\bf D67}
  (2003) 073006, [\href{http://xxx.lanl.gov/abs/hep-ph/0212229}{{\tt
  hep-ph/0212229}}]. [Erratum-ibid.D73:119901,2006].

\bibitem{Braun:2002en}
V.~M. Braun, S.~Gottwald, D.~Y. Ivanov, A.~Sch{\"a}fer, and L.~Szymanowski {\em
  Phys. Rev. Lett.} {\bf 89} (2002) 172001,
  [\href{http://xxx.lanl.gov/abs/hep-ph/0206305}{{\tt hep-ph/0206305}}].

\bibitem{Pire:2009nn}
B.~Pire and L.~Szymanowski \href{http://xxx.lanl.gov/abs/0909.0098}{{\tt
  arXiv:0909.0098}}.

\bibitem{Nyffeler:2009uw}
A.~Nyffeler {\em PoS} {\bf CD09} (2009) 080,
  [\href{http://xxx.lanl.gov/abs/0912.1441}{{\tt arXiv:0912.1441}}].

\bibitem{Belyaev:1984ic}
V.~Belyaev and Y.~Kogan {\em Yad. Fiz.} {\bf 40} (1984) 1035.

\bibitem{Balitsky:1985aq}
I.~Balitsky, A.~Kolesnichenko, and A.~Yung {\em Sov. J. Nucl. Phys.} {\bf 41}
  (1985) 178.

\bibitem{Ball:2002ps}
P.~Ball, V.~M. Braun, and N.~Kivel {\em Nucl. Phys.} {\bf B649} (2003) 263,
  [\href{http://xxx.lanl.gov/abs/hep-ph/0207307}{{\tt hep-ph/0207307}}].

\bibitem{Bergman:2008sg}
O.~Bergman, G.~Lifschytz, and M.~Lippert {\em JHEP} {\bf 0805} (2008) 007,
  [\href{http://xxx.lanl.gov/abs/0802.3720}{{\tt arXiv:0802.3720}}].

\bibitem{Gorsky:2009ma}
A.~Gorsky and A.~Krikun {\em Phys. Rev.} {\bf D79} (2009) 086015,
  [\href{http://xxx.lanl.gov/abs/0902.1832}{{\tt arXiv:0902.1832}}].

\bibitem{Vainshtein:2002nv}
A.~Vainshtein {\em Phys. Lett.} {\bf B569} (2003) 187,
  [\href{http://xxx.lanl.gov/abs/hep-ph/0212231}{{\tt hep-ph/0212231}}].

\bibitem{Kim:2004hd}
H.-C. Kim, M.~Musakhanov, and M.~Siddikov {\em Phys. Lett.} {\bf B608} (2005)
  95--106, [\href{http://xxx.lanl.gov/abs/hep-ph/0411181}{{\tt
  hep-ph/0411181}}].

\bibitem{Dorokhov:2005pg}
A.~E. Dorokhov {\em Eur. Phys. J.} {\bf C42} (2005) 309,
  [\href{http://xxx.lanl.gov/abs/hep-ph/0505007}{{\tt hep-ph/0505007}}].

\bibitem{Goeke:2007nc}
K.~Goeke, H.-C. Kim, M.~Musakhanov, and M.~Siddikov {\em Phys. Rev.} {\bf D76}
  (2007) 116007, [\href{http://xxx.lanl.gov/abs/0708.3526}{{\tt
  arXiv:0708.3526}}].

\bibitem{Nam:2008ff}
S.-i. Nam, H.-Y. Ryu, M.~Musakhanov, and H.-C. Kim {\em J. Korean Phys. Soc.}
  {\bf 55} (2009) 429--434, [\href{http://xxx.lanl.gov/abs/0804.0056}{{\tt
  arXiv:0804.0056}}].

\bibitem{Ioffe:2009yi}
B.~Ioffe {\em Phys. Lett.} {\bf B678} (2009) 512--515,
  [\href{http://xxx.lanl.gov/abs/0906.0283}{{\tt arXiv:0906.0283}}].

\bibitem{Frasca:2011zn}
M.~Frasca and M.~Ruggieri {\em Phys. Rev.} {\bf D83} (2011) 094024,
  [\href{http://xxx.lanl.gov/abs/1103.1194}{{\tt arXiv:1103.1194}}].

\bibitem{Buividovich:2009ih}
P.~Buividovich, M.~Chernodub, E.~Luschevskaya, and M.~Polikarpov {\em Nucl.
  Phys.} {\bf B826} (2010) 313, [\href{http://xxx.lanl.gov/abs/0906.0488}{{\tt
  arXiv:0906.0488}}].

\bibitem{Dunne:2004nc}
G.~V. Dunne \href{http://xxx.lanl.gov/abs/hep-th/0406216}{{\tt
  hep-th/0406216}}.

\bibitem{Endrodi:2013cs}
G.~Endr\H{o}di \href{http://xxx.lanl.gov/abs/1301.1307}{{\tt arXiv:1301.1307}}.

\bibitem{Colangelo:2003hf}
G.~Colangelo and S.~{D\"urr} {\em Eur. Phys. J.} {\bf C33} (2004) 543,
  [\href{http://xxx.lanl.gov/abs/hep-lat/0311023}{{\tt hep-lat/0311023}}].

\bibitem{Gatto:2010pt}
R.~Gatto and M.~Ruggieri {\em Phys. Rev.} {\bf D83} (2011) 034016,
  [\href{http://xxx.lanl.gov/abs/1012.1291}{{\tt arXiv:1012.1291}}].

\bibitem{Cohen:2007bt}
T.~D. Cohen, D.~A. McGady, and E.~S. Werbos {\em Phys. Rev.} {\bf C76} (2007)
  055201, [\href{http://xxx.lanl.gov/abs/0706.3208}{{\tt arXiv:0706.3208}}].

\bibitem{Andersen:2012dz}
J.~O. Andersen {\em Phys. Rev.} {\bf D86} (2012) 025020,
  [\href{http://xxx.lanl.gov/abs/1202.2051}{{\tt arXiv:1202.2051}}].

\bibitem{Chernodub:2010qx}
M.~Chernodub {\em Phys. Rev.} {\bf D82} (2010) 085011,
  [\href{http://xxx.lanl.gov/abs/1008.1055}{{\tt arXiv:1008.1055}}].

\bibitem{Bali:2011uf}
G.~Bali, F.~Bruckmann, G.~Endr\H{o}di, Z.~Fodor, S.~Katz, {\em et.~al.} {\em
  PoS} {\bf LATTICE2011} (2011) 192,
  [\href{http://xxx.lanl.gov/abs/1111.5155}{{\tt arXiv:1111.5155}}].

\bibitem{Fukushima:2012kc}
K.~Fukushima and Y.~Hidaka \href{http://xxx.lanl.gov/abs/1209.1319}{{\tt
  arXiv:1209.1319}}.

\bibitem{Nielsen:1980sx}
N.~Nielsen {\em Am. J. Phys.} {\bf 49} (1981) 1171.

\bibitem{Wilczek:1996bw}
F.~Wilczek \href{http://xxx.lanl.gov/abs/hep-th/9609099}{{\tt hep-th/9609099}}.

\bibitem{Grozin:2008yd}
A.~Grozin \href{http://xxx.lanl.gov/abs/0803.2589}{{\tt arXiv:0803.2589}}.

\bibitem{Durr:2010vn}
S.~{D\"urr}, Z.~Fodor, C.~Hoelbling, S.~Katz, S.~Krieg, {\em et.~al.} {\em
  Phys. Lett.} {\bf B701} (2011) 265,
  [\href{http://xxx.lanl.gov/abs/1011.2403}{{\tt arXiv:1011.2403}}].

\bibitem{Jamin:2002ev}
M.~Jamin {\em Phys. Lett.} {\bf B538} (2002) 71,
  [\href{http://xxx.lanl.gov/abs/hep-ph/0201174}{{\tt hep-ph/0201174}}].

\end{thebibliography}\endgroup

\end{document}